\documentclass{pasj00}
\draft

\begin{document}
\SetRunningHead{K. Hamaguchi, M. F. Corcoran, and K. Imanishi}{Chandra observations of a young magnetic B star}
\Received{2002/10/31}
\Accepted{2003/8/19}

\title{Chandra Observations of a Young Embedded Magnetic B Star in
the $\rho$ Ophiuchus Cloud}

\author{Kenji \textsc{Hamaguchi},$^{1,2}$ Michael F. \textsc{Corcoran},$^{1,3}$ and Kensuke \textsc{Imanishi}$^{4}$}
\affil{$^{1}$Laboratory for High Energy Astrophysics, Goddard Space Flight Center,
Greenbelt, MD 20771, USA}
\email{kenji@milkyway.gsfc.nasa.gov,corcoran@barnegat.gsfc.nasa.gov}
\affil{$^{2}$National Research Council, 500 Fifth Street, NW, Washington, DC 20001, USA}
\affil{$^{3}$Universities Space Research Association, 7501 Forbes 
Blvd, Ste 206, Seabrook, MD 20706, USA}
\affil{$^{4}$Department of Physics, Faculty of Science, Kyoto University\\
Kitashirakawa-oiwakecho, Sakyo-ku, Kyoto 606-8502}
\email{kensuke@cr.scphys.kyoto-u.ac.jp}
\maketitle


\begin{abstract}
This paper reports on an analysis of two Chandra X-ray observations of
the young magnetic B star $\rho$ Ophiuchus S1.
X-ray emission from the star was detected in both observations.
The average flux was almost the same in both,
but during each observation the flux showed significant time
variations by a factor of two on timescales of 20--40 ks.
Each spectrum could be fit by either an absorbed power-law model with a
photon index of $\sim3$
or a thin-thermal plasma model with a temperature of $\sim2$ keV and
an extremely low
metal abundance ($\lesssim0.1$ solar).
The spectrum of the first observation has a weak-line
feature at about 6.8 keV, which might correspond to
highly ionized iron K$\alpha$.
In contrast, the spectrum of the second observation apparently shows
a weak edge absorption component at $E\sim4$ keV.
The continuum emission and $\log (L_{\rm X}/L_{{\rm bol}})\sim-6$ are
similar to those of young intermediate-mass stars (Herbig Ae/Be stars),
although the presence of a strong magnetic field (inferred from the
detection of non-thermal radio emission)
has drawn an analogy between $\rho$ Ophiuchus S1 and magnetic chemically
peculiar (MCP) stars.
If the X-ray emission is thermal,
the small abundances that we derived might be related to the inverse first-ionization potential (FIP) effect,
though there is no significant trend as a function of FIP from our model fits.
If the emission is non-thermal, it might be produced by high-energy
electrons in the magnetosphere.
\end{abstract}

\section{Introduction}
Intermediate-mass stars (1.5 $M_{\odot}\lesssim M \lesssim$8 $M_{\odot}$) do not generally
exhibit magnetic activity.
This is explained by the absence of a surface convection zone to
generate a solar-type dynamo to amplify the magnetic field.
Certain populations of intermediate-mass stars, however, are thought to have
magnetic fields.
Herbig Ae/Be (HAeBe) stars are pre-main sequence intermediate-mass stars, some of which are thought to possess significant magnetic fields (e.g., Catala et al. 1993).  These fields may be fossil remnants from the parent molecular cloud amplified by the stellar accretion process (e.g., Moss 2001). Magnetic Ap/Bp stars, also called magnetic chemically peculiar (MCP)
stars, are intermediate-mass main sequence (MS) stars (Hubrig et al.
2000), which exhibit strong Zeeman effects in their absorption lines,
implying the presence of dipole magnetic fields of a few
hundred to several thousand Gauss (Borra et al. 1982).
The magnetic fields of MCP stars may be the fossil remains of
fields present in the earlier HAeBe phase.

Both MCP and HAeBe stars are X-ray sources.
MCP stars typically have $\log (L_{\rm X}/L_{\rm bol})<-6$ (Drake et al. 1994).
Among the limited sample of MCP stars with X-ray spectra, the derived plasma 
temperatures are typically less than 1 keV (Babel, Montmerle 1997, hereafter BM97; 
Bergh\"{o}fer et al. 1996).
The $\log (L_{\rm X}/L_{\rm bol}$) ratio of HAeBe stars can reach $-4$,
with observed plasma temperatures near 2 keV
(Zinnecker, Preibisch 1994; Skinner, Yamauchi 1996; Yamauchi et 
al. 1998; Hamaguchi et al. 2000; Hamaguchi 2001).
However, X-ray emission, especially from MCP stars, does not have a clear correlation with the
stellar parameters (Drake 1998),
so that it has been argued that the X-rays could arise from hidden 
low-mass companions.

Different X-ray emission mechanisms have been proposed for HAeBe and MCP stars.  X-ray emission from MCP stars is thought to arise from the collision of magnetically confined wind plasma in a closed 
magnetosphere
(Havnes, Goertz 1984; BM97), while in HAeBe stars X-ray emission is thought to be produced by  magnetic dynamo activity related to mass accretion
(e.g., Hamaguchi 2001).
However, these two classes of stars have similar magnetic field characteristics, so 
the observed differences in their X-ray properties might be due to the
change in the circumstellar properties (e.g., mass accretion rate, circumstellar disks) with age, as is suggested by Hamaguchi (2001).

$\rho$ Ophiuchus S1 (hereafter S1) is one of the
best examples of stars whose evolutionary phase is thought to be between the 
HAeBe and MCP phase.
It is a B3 V star associated with the $\rho$ Ophiuchus cloud core A,
and has the following derived stellar parameters: distance 
$d \sim$120 pc, effective temperature $T_{\rm eff}\sim$16000 K,
bolometric luminosity $L_{\rm bol}\sim$1100 $L_{\odot}$,
and radius $r_{\ast} \sim$3$\times$10$^{11}$ cm
(Andr{\' e} et al. 1988, hereafter A88; Knude, H$\phi$g 1998;
N\"urnberger et al. 1998).
Its large visual extinction ({\it A$_{V}$} = 11$^{\rm m}$.6) does
not allow the detection of emission lines needed to classify it as an MCP star,
but the detection of polarized non-thermal radio emission
(A88), which probably comes from gyro-synchrotron
particles in a large magnetosphere, suggests that S1 possesses a strong magnetic field.
S1 has many characteristics of youth; 
a class III object with a double-peaked spectral energy distribution (e.g., Ward-Thompson 1993; Wilking et al. 2001),
possession of a compact H~{\sc ii} region (A88) and proximity to a plausible star 
forming cloud SM1 (Motte et al. 1998).
S1 may have dissipated most of its disk ($<2.3\times
10^{-3}M_{\odot}$, N{\"u}rnberger et al. 1998),
so that it should be near or already on the MS.

Observations of S1 with Einstein, ROSAT and ASCA
have shown relatively strong X-ray
emission for B stars ($L_{\rm X} \approx 10^{30-31}$ erg s$^{-1}$,
Montmerle et al. 1983; Casanova et al. 1995; Kamata et al. 1997), but
those observations did not derive timing and spectral
properties due to limited photon statistics and
severe contamination by a nearby source.
This paper compares two Chandra observations of the star to attempt to characterize the X-ray emission properties and to constrain the 
emission mechanism.
A brief summary of the X-ray time variability and the spectral parameters of S1 in one of the observations was previously given in Gagn{\' e} (2001) and Skinner, Daniel, and 
Gagn{\' e} (2002).

\section{Observations and Data Reduction}
S1 was observed twice with the Chandra X-ray observatory in the
timed event mode with the Advanced CCD Imaging Spectrometer (ACIS, Weisskopf et al. 2002).
The first observation (Obs1) was a 100 ks exposure made by the imaging array (ACIS-I) on 2000 April 13.
The telescope optical axis on the ACIS-I array pointed at the $\rho$
Ophiuchus cloud core F
($\alpha_{2000}$ = $16^{\rm h}27^{\rm m}$18\fs1, $\delta_{2000}$ =
$-24^{\circ}34'21\farcs9$, Loren et al. 1990). In this observations
S1 was 14.$'$8 off-axis, and was detected on the ACIS-S3 chip.
The second observation (Obs2) was a 96 ks exposure with ACIS-I made on 2000 May 15.  The exposure was centered on 
the $\rho$ Ophiuchus A cloud ($\alpha_{2000}$ =
$16^{\rm h}26^{\rm m}$35\fs3, $\delta_{2000}$ =$-24^{\circ}23'12\farcs9$).
S1 was $0.'3$ off-axis on the ACIS-I3 chip.
For each observation we utilized the level-2 screened event data, which were
processed at the Chandra X-ray Center (CXC)
(processing software, ver. R4CU5UPD13.2 for Obs1, ver. R4CU5UPD13.2 for Obs2).
Post-production data reduction and further analyses were performed with the software
packages CIAO 2.1.3 and FTOOLS 4.2.

\section{Analysis and Results}
\subsection{Source Detection and Event Extraction}
In each observation, a bright X-ray source was detected at the
optical position of S1
(error circle, Obs1: $\sim$15\arcsec, Obs2: $\sim$0\farcs5) by using the {\it
wavdetect} package (figure \ref{fig:image}).
In Obs2, a circle of radius of $\sim$1\farcs3 included 
95\% of the source photons, while in Obs1 the 95\% radius was 
$\sim35''$ because of the large off-axis angle.
In the Obs2 image, no other X-ray source was detected within 
the 95\% radius circle of Obs1.
The X-rays in both observations should therefore have come from the same
source uncontaminated by nearby sources. 
The coordinates derived from the satellite attitude data
have a small systematic
offset,\footnote{$\langle$http://cxc.harvard.edu/cal/ASPECT/celmon/$\rangle$.} which 
we corrected by a cross-correlation of
the Chandra detected sources for the Obs2 data with near-infrared counterparts in the 2MASS point 
source catalog\footnote{$\langle$http://www.ipac.caltech.edu/2mass/releases/second/doc/$\rangle$.} [($\Delta\alpha$, $\Delta\delta$) = ({0\farcs0, 1\farcs0})].
After the correction, the position of the X-ray source is
($\alpha_{2000}$, $\delta_{2000}$) = ($16^{\rm h} 26^{\rm m} 34\fs21$,
$-24^{\circ} 23' 28\farcs2$).
The 2MASS position of S1 is 0\farcs27 distant and the radio 
position of S1 is 0\farcs57 distant
so that these positions are within the X-ray error circle.
The X-ray source thus corresponds to the position of S1.

We extracted source events from a circular
region centered on the X-ray position, with a radius larger than the radius of the 95\%
circle to gather all X-ray events.
Background events were extracted from source-free regions
(Obs1: 19 arcmin$^{2}$, Obs2: 59 arcmin$^{2}$, see 
figure \ref{fig:image}).
The ratio of normalized background to source counts between 
0.5{\em --}9 keV is not negligible in Obs1 (18.4\%), but it is quite small in Obs2 (0.5\%)
because of its small source region.

\subsection{Timing Analysis}
\label{subsec:timingana}
Figure  \ref{fig:curve} shows the background-subtracted light curves
(left: Obs1, right: Obs2)
in the total (0.5{\em --}9 keV), soft (0.5{\em --}2 keV) and hard (2{\em --}9 keV)
bands.
Both total-band light curves show significant variations,
and neither are consistent with a constant-flux model (table \ref{tal:varexp}).
However the background levels in both are almost constant.
In Obs1, the light curve gradually decreases with a small flux increase near
the middle of the observation ($t \sim5\times10^{4}$ s).
The light curve can be reproduced by a constant plus exponential decay
($e$-folding time $\sim$40 ks, table \ref{tal:varexp}).
This variability is also seen in both the soft and hard bands.
In contrast, the flux in Obs2 is almost constant, but
then increases abruptly to 0.04 count s$^{-1}$ at $t$ = 80 ks.
The total band light curve at 80 $< t < $ 100 ks can be fit by a linear model 
with a slope of 1.0$^{+0.4}_{-0.3}$
$\times10^{-6}$ count s$^{-2}$ ($\chi^2$/d.o.f = 5.9/8),
which corresponds to a variation time scale of $\sim$20 ks.
The standard deviations are $3.8\times10^{-3}$ count s$^{-1}$ (0.5{\em --}2 keV) and
$4.7\times10^{-3}$ count s$^{-1}$ (2{\em --}9 keV) in Obs1, and 
$2.9\times10^{-3}$ count s$^{-1}$
(0.5{\em --}2 keV) and $4\times10^{-3}$ count s$^{-1}$ (2{\em --}9 keV) in Obs2,
respectively.
The standard deviations are somewhat larger in the hard band in each observation.

\subsection{Spectral Analysis}
The time-averaged spectra of S1 in Obs1 and Obs2 are shown in figures
\ref{fig:spec1stmekawabs}  and \ref{fig:spec2ndmekawabs}, respectively.
Response matrices and ancillary response function tables at the source position for both spectra
were generated by the CIAO 2.1.3 ``mkrmf'' and ``mkarf'' commands.
Although other bright $\rho$ Ophiuchus X-ray sources have thermal
X-rays (Imanishi et al. 2001),
S1 does not show any features in its spectra except for
a marginal hump at $\sim$6.5 keV in Obs1, and a weak edge feature in Obs2.
We attempted to fit the spectra with an absorbed thin-thermal plasma model
(MEKAL code, Mewe et al. 1985, 1986; Kaastra 1992; Liedahl et al.  1995) and an absorbed power-law model.
The spectrum in Obs1 can then be reproduced with either a very low
metal abundance plasma ($Z \lesssim$0.1 solar) or
an absorbed power-law model with a steep photon index ($\Gamma \sim$3.4)
(table \ref{tbl:1stspec}).
We can see an emission feature at around 6.5 keV, and 
therefore added a narrow Gaussian component near 6.5 keV 
to the thermal model
(table \ref{tbl:1stspec}).
The best-fit line center energy is $\sim6.8\pm0.2$ keV, consistent with either He-like 
(6.7 keV) or H-like (6.9 keV) iron,
but inconsistent with neutral iron (6.4 keV).
Including a Gaussian line with a power law model yields the same line energy.
The line intensity can be fit by assuming a 1{\em --}2 keV plasma with solar iron abundance and 
$N_{\rm H} \approx  3\times10^{22}$ cm$^{-2}$, 
but the model also requires a small metal abundance of $\sim$0.1 solar for other elements.

On the other hand, the spectrum in Obs2 rejects an absorbed  single temperature model
at the 96\% confidence level
($\chi^2$/d.o.f = 129.3/102 for a thermal model) due to
a deficit in the flux near 4 keV and an excess in flux above 4 keV (table \ref{tbl:2ndspec} and figure \ref{fig:spec2ndmekawabs}).
These residuals are not due to background, since the area-normalized background level 
is below $\sim$10$^{-5}$ count s$^{-1}$ keV$^{-1}$.
These residuals can be fit by including an absorption edge at $E \sim4$ keV. 
A thermal model  including an edge feature reduces the $\chi^2$ value
to an acceptable range ($\chi^2$/d.o.f = 116.5/100; figure 
\ref{fig:spec2ndmekaedgewabs}).
The spectral parameters, except for the plasma temperature, are almost
the same as those in Obs1.
The column density $N_{\rm H}$ is $\sim$2$\times$10$^{22}$ cm$^{-2}$,
consistent with the $V$-band extinction of S1 ($A_{V}\sim$11$^{\rm m}$.7),
using the $N_{\rm H}${\em --}$A_{V}$ relation appropriate for  the $\rho$
Ophiuchus cloud (Imanishi et al. 2001).
The metal abundance is quite low so that, like Obs1, 
the spectrum for Obs2 can also be fit by an absorbed power-law model
(if an edge component is included).
As far as we are aware, an edge feature, like that seen in the  Obs2 spectrum, has never been 
seen in any other stellar X-ray spectra.
We do not think that this feature is an instrumental effect: the data do not suffer a severe event pile-up, nor do nearby 
sources show any similar edge feature.
Neither a two-temperature model nor the addition of a Gaussian line at 3.65 keV could reproduce the apparent edge feature.

If this edge is real, its observed threshold energy (3.84 keV $<E_{\rm edge} < 4.09$ keV)
includes the K-shell binding energies of abundant elements Ar and Ca in neutral 
or ionized states (Lotz 1968).
For Ca, the energy of the edge only includes Ca~{\sc i} (4.041 keV) and Ca~{\sc ii} (4.075 keV), which exist $\lesssim 10^{4}$ K
(Arnaud, Rothenflug 1985).
We refit the spectrum allowing for the Ca abundance in absorption to vary,
constraining the abundance of other elements at their solar values.
For either the thermal or non-thermal model,
Ca in the absorber would need to have an abundance of $\sim$500 (180--3800) solar to reproduce the observed edge, which seems to be unreasonably high.
For Ar, the edge energy is consistent with Ar~{\sc xv} (3.887 keV){\em --}~{\sc xvi} (3.953 keV), which
mainly exist in the temperature range $6.5< \log T$(K)$ < 6.7$ (Arnaud, Rothenflug 1985).
We simulated a ``warm absorber'' model for Ar at $\log T$(K)$\sim6.6$
by multiplying several edge components corresponding to the Ar~{\sc xiii}{\em --}{\sc xvii} states.
We could reproduce the edge with an optical depth of
Ar~{\sc xvii} of 0.22 (0.11--0.32),
equivalent to $N_{\rm H}^{\rm abs} \sim10^{24}/Z_{\rm Ar}^{\rm abs}$ cm$^{-2}$, where
$N_{\rm H}^{\rm abs}$ is the hydrogen column density and $Z_{\rm Ar}^{\rm abs}$ the Ar abundance 
of the warm absorber.
Assuming that the plasma has a scale height of $\sim 1$ stellar radius ($3\times10^{11}$ cm) and a density similar to that of the inner part of the
magnetosphere ($10^{12}$cm$^{-3}$, Havnes, Goertz 1984),
the Ar abundance should be $\sim$3 solar.

\section{Discussion}

\subsection{General Characteristics of the X-Ray Emission}
\label{subsec:genechara}
The X-ray properties of S1, namely its relatively hard emission and its X-ray variability,
are more similar to those of low-mass young stars than of early-type MS stars.
Certainly, S1 has a faint close companion ($K = 8^{\rm m}$)
at a projected separation of $\sim$0\farcs02,
whose spectral type is unknown (Simon et al. 1995).
It is thus possible that some or all of the observed X-ray emission might be produced by this 
low-mass companion star.
However, according to the ROSAT survey of the Taurus cloud (Neuh{\"a}user et al. 1995),
more than 90\% of optically selected low-mass stars have
X-ray luminosities less than 10$^{30}$ erg s$^{-1}$.
Since the X-ray luminosity of S1 is above $10^{30}$ erg s$^{-1}$,
it is likely that most of the observed X-ray emission comes from S1 itself.

The Chandra spectra show that S1 has $-6.5<\log (L_{\rm X}/L_{\rm bol})<-5.5$ in the {\it ROSAT} band (0.1--2.4 keV).
The $\log (L_{\rm X}/L_{\rm bol})$ ratio is larger than that of He-rich Bp stars with strong magnetic fields 
[$\log (L_{\rm X}/L_{\rm bol}) \sim-7$], and is closer to that of non-magnetic Bp stars or Ap stars
[$\log (L_{\rm X}/L_{\rm bol}) <-6$, Drake et al. 1994].
In contrast, the $L_{\rm X}/L_{\rm bol}$ ratio of S1 is within the 
range of that of HAeBe stars [$\log (L_{\rm X}/L_{\rm bol}) <-$4,
Zinnecker, Preibisch 1994; Hamaguchi 2001].
On the other hand,
the plasma temperature of S1 ($kT \sim 2$ keV) is larger
than that of MCP stars measured with ROSAT ($kT \lesssim1$ keV, e.g. BM97 though they
note a hint of a hot component of $\sim$4.5 keV on the Ap star IQ Aur, too),
but is typical of temperatures of 
HAeBe and young MS stars (Hamaguchi 2001; Feigelson et al. 2002).
Thus, the X-ray properties of S1 seem to be closer to those of  HAeBe stars than to  
the more-evolved MCP stars.

\subsection{X-Ray Emission Mechanisms}
The lack of significant X-ray line emission seems to be consistent with either emission from a thermal plasma with non-solar abundances or a non-thermal source. In principle, the emission could be a composite of thermal and non-thermal emission, but for simplicity we consider each separately.  

\subsubsection{Thermal emission from magnetically-confined plasma}
\label{subsec:casethermal}
Thermal emission could arise from magnetic heating of gas within the stellar magnetosphere or from wind-shocked gas.  In a simulation of wind-shocked gas by BM97,
the derived plasma temperatures are less than 1 keV for a plausible range of physical parameters 
(see table 3 in BM97).  This is lower than the observed temperature ($kT \sim 2$ keV) of S1, though we point out that the simulations of BM97 assume a star of spectral type A0.

The observed emission from S1 does, however, share some characteristics with flare heated plasma. 
The X-ray variability in both Obs1 and Obs2 is less significant in the 
soft band than in the hard band 
(see subsection \ref{subsec:timingana}),
which is a property seen in the thermal emission from flares in low-mass stars.
At {\it kT} $\sim$2 keV,
the shock propagation speed, $v_{\rm prop}$, would be $\sim$(5{\em --}10)$\times10^7$ cm s$^{-1}$ if shock propagates at the sound speed. 
Because the X-ray variation time scale  is $\Delta t  \sim 20$ ks in Obs2,
the plasma scale ($l_{\rm em} < v_{\rm prop} \Delta t$) is less
than $\sim2\times10^{12}$ cm ($\sim7 r_{\ast}$),
which implies a plasma density, $n_{\rm em} \sim EM/l_{em}^{3}> 2\times10^{8}$ cm$^{-3}$, where 
the emission measure $EM  \sim 2\times10^{53}$ cm$^{-3}$.
The plasma scale $l_{\rm em}$ that we derive is smaller than the size of the closed magnetosphere
($d \sim$12.8 $r_{\ast}$, Andr{\'e} et al. 1991), 
and the derived density in the X-ray plasma is similar to the gas density within several stellar radii of MCP stars (Havnes, Goertz, 1984).

Because S1 is near to the Sun, 
the global abundances of S1 are expected to be near the solar value.  
A thermal plasma model for the emission, however, requires sub-solar abundances, which if real 
might indicate some elemental selection mechanism.
Stellar coronae sometime show abundance anomalies that
depend on first ionization potentials (FIP) (G\"udel et al. 2001).
To test whether a similar mechanism is at work here, 
we estimated the upper limits to the abundances of Mg, Si, S, Ar, Ca, and Fe from the spectrum in Obs1.
We consider two typical cases, where the abundances of He, C, N, and O are 1 solar and 0.3 solar
(typical values observed in stellar X-rays, e.g. Kitamoto, Mukai 1996).
The results are given in table \ref{tbl:lineuppplimit}.  
There is no significant trend as a function of FIP from our model fits.  
This suggests that elemental abundances are basically small for all elements
irrespective of their FIP values.

\subsubsection{Non-thermal emission}
\label{subsec:casenonthermal}
S1 is a non-thermal radio source, which implies that a significant population of
gyro-synchrotron electrons are associated with the star (A88).
Linearly extrapolating the radio spectrum to the X-ray band (see figure 4 in A88), however, 
yields an X-ray flux well below the observed X-ray emission level (figure  \ref{fig:wrspow}).
Thus, the X-ray emission is not explained by the same gyro-synchrotron electron population.
Observable gyro-synchrotron X-rays from S1 require 10 GeV electrons plus a field of 
a few hundred gauss for the synchrotron process, but
the radio emission only suggests the presence of MeV electrons around S1 (A88).
However, MeV electrons could upscatter stellar UV photons to X-ray energies by the inverse-Compton process  (a similar process was considered for producing hard X-ray tails of massive MS stars, Chen, White 1991).

On the other hand, high-energy electrons which hit a dense region, such as the stellar surface,
could produce observable non-thermal bremsstrahlung X-rays.
In solar flares, matter accelerated by the reconnection of magnetic loops above the solar surface
falls to the surface with $v \sim$ 3000 km s$^{-1}$, which produces non-thermal emission, which is dominant above 20 keV
(Sakao et al. 1998).
If matter infalls on S1 with a slower infall velocity
(for example, free fall velocity of $\sim$600 km s$^{-1}$), 
the thermal component is cooler, and
bremsstrahlung X-rays from non-thermal electrons could conceivably be observable in the Chandra band.

\section{Summary}
The magnetic B star S1 is identified as an X-ray source with a
large X-ray luminosity (log $L_{\rm X} \sim$ 30.3 erg s$^{-1}$)
with a precision of $\sim$ 0\farcs5.
The observations give good supporting evidence that the intermediate-mass star S1 itself
 is a source of the  X-ray emission.
The X-rays do not show the characteristics of X-ray emission from normal early-type MS stars nor MCP stars, 
but are more similar to those characteristics of HAeBe stars; 
S1 shows $\log (L_{\rm X}$/$L_{\rm bol})\sim- 6$, with small but significant 
X-ray time variations and 
significantly hard X-ray emission, corresponding to $kT \sim2$ keV.
The X-ray emission of S1 might be related to its youth.
The spectra do not show strong emission lines, suggesting either anomalously low abundances (0.1 solar),
possibly caused by selective abundance reductions, if the emitting plasma is thermal, 
or the presence of significant populations of non-thermal electrons, if the emission is non-thermal.
The X-ray emission mechanisms might be related to the non-thermal radio emission of S1.
In order to address the X-ray emission mechanism,
we have to determine sensitive upper limits for each emission line and
confirm the presence of the edge feature.
Deeper observations by XMM-Newton, Chandra and high resolution spectroscopy
with Astro-E II will help to address these problems. 

We greatly appreciate useful discussions with S. Drake and K. Koyama, 
and comments by D. Davis, K. Gendreau, K. Kikuchi, K. Motohara, and R. Mushotzky. This work was performed while the authors held a National Research Council Research Associateship Award
at NASA/GSFC and awards by National Space Development Agency of Japan (NASDA) and 
the Japan Society for the Promotion of Science for Young Scientists (JSPS).

\clearpage

\begin{table}
\begin{center}
\caption{Fitting results of the light curves.}
\label{tal:varexp}
\begin{tabular}{llllll}  \hline\hline
Observations&&\multicolumn{2}{c}{Obs1}&& Obs2 \\ \cline{3-4}\cline{6-6}
Model&&Cons.&Cons.~+~Exp.&&Cons. \\ \hline
Mean &[10$^{-2}$ count s$^{-1}$]  &2.1&1.4 (0.5--1.8)&&2.6\\
$e$-folding time &[10$^4$ s] &...&3.8 (2.2--8.3)&&...\\ \hline
$\chi^2$/d.o.f.		   &&154.9/51&\multicolumn{1}{l}{49.7/49}&&95.9/49 \\ \hline
\multicolumn{5}{@{}l@{}}{\hbox to 0pt{\parbox{85mm}{\footnotesize
       \par\noindent
	Cons.: Constant model, Exp.: Exponential model
       \par\noindent
	The numbers in parentheses are the 90\% confidence intervals.
}\hss}}
\end{tabular}
\end{center}
\end{table}

\begin{table}
\begin{center}
\caption{Fitting results of the spectrum in the 1st observation (Obs1).}
\label{tbl:1stspec}
\begin{tabular}{lllll}  \hline\hline
Model&&1T&1T~+~gauss&power\\ \hline
{\it N$_{\rm H}$}&[10$^{22}$ cm$^{-2}$]& 1.9 (1.7{\em --}2.1)&1.9 (1.7{\em --}2.1)    &2.4 (2.1{\em --}2.7) \\
{\it kT}/$\Gamma$&[keV]/ & 1.7 (1.4{\em --}1.9)        &1.6 (1.4{\em --}1.9)           &3.4 (3.1{\em --}3.6) \\
Abundance&[solar]    & 0.0 (0.0{\em --}0.11)   &0.0 (0.0{\em --}0.09)                    &...         \\
Emission measure&[10$^{53}$ cm$^{-3}$]& 2.8 (2.2{\em --}3.4)   &2.8 (2.2{\em --}3.5)     &...         \\
Line center             &[keV]&...&6.77 (6.56{\em --}6.99)\footnotemark[*]   &...\\
Line flux&[10$^{-6}$ photon cm$^{-2}$ s$^{-1}$]&...&3.1 (1.1{\em --}5.0)          &...         \\
{\it L$_{\rm X}$} (0.5{\em --}10 keV)\footnotemark[**]&[10$^{30}$ erg s$^{-1}$]&1.8      &1.9       
&4.8\\ \hline 
$\chi^2$/d.o.f             &&118.5/106          &111.5/104                 &116.5/107   \\ 
\hline
   \multicolumn{5}{@{}l@{}}{\hbox to 0pt{\parbox{150mm}{\footnotesize
       \par\noindent
        The errors listed in parenthesis quote for 90\% confidence.
       \par\noindent
       Line width ($\sigma$) of the Gaussian component is fixed on zero.
       \par\noindent
       \footnotemark[$*$] Other parameters, except for the line normalization, are temporarily 
frozen in the error estimate.
       \par\noindent
       \footnotemark[$**$] Absorption corrected X-ray luminosity assuming the distance of 120 pc.
     }\hss}}
\end{tabular}
\end{center}
\end{table}

\begin{table}
\begin{center}
\caption{Fitting results of the spectrum in the 2nd observation (Obs2).}
\label{tbl:2ndspec}
\begin{tabular}{llllll}  \hline\hline
Model&&1T&1T~$\times$~edge&power&power~$\times$~edge \\ \hline
{\it N$_{\rm H}$}&[10$^{22}$ cm$^{-2}$]& 2.0    & 1.8 (1.6{\em --}2.0)   & 2.5      &2.2 (2.1{\em --}2.4)
 \\
{\it kT}/$\Gamma$&[keV]/ & 1.9  & 2.5 (2.1{\em --}3.1)   &3.2       &2.7 (2.5{\em --}3.0)\\
Abundance &[solar]    & 0.21    & 0.14 (0.0{\em --}0.28) &...       &... \\
Emission measure &[10$^{53}$ cm$^{-3}$]&...             & 1.6 (1.3{\em --}1.9)           &...     
  &...           \\     
Threshold energy\footnotemark[*]&[keV]             & ...        & 3.96 (3.84{\em --}4.07)& ...    
  &4.00 (3.89{\em --}4.09) \\
Absorption depth\footnotemark[**]&         & ...        & 0.53 (0.28{\em --}0.81)& ...      &0.67 
(0.48{\em --}0.90) \\
{\it L$_{\rm X}$} (0.5{\em --}10 keV)\footnotemark[***]&[10$^{30}$ erg s$^{-1}$]&...     &1.5       
          &...       &2.6 \\ \hline 
$\chi^2$/d.o.f             && 129.3/102 & 116.5/100       & 132.3/103&111.3/101 \\ \hline
   \multicolumn{6}{@{}l@{}}{\hbox to 0pt{\parbox{150mm}{\footnotesize
       \par\noindent
        The errors listed in parenthesis quote for 90\% confidence.
       \par\noindent
       \footnotemark[$*$] Threshold energy of the edge component.
       \par\noindent
       \footnotemark[$**$] Absorption depth at the threshold of the edge component.
       \par\noindent
       \footnotemark[$***$] Absorption-corrected X-ray luminosity assuming the distance of 
120 pc.
     }\hss}}
\end{tabular}
\end{center}
\end{table}

\begin{table}
\begin{center}
\caption{Emission line upper-limit (Obs1).}
\label{tbl:lineuppplimit}
\begin{tabular}{cccc}  \hline\hline
& FIP & \multicolumn{2}{c}{Metal abundance} \\
& [eV]  & [solar] & [solar]\\ \hline
He, C, N, O&   & 0.3 & 1\\ \hline
Ca & 6.1 &$<$0.37 & $<$0.53\\
Mg & 7.6 &0.01{\em --}0.87 & 0.11{\em --}1.37\\
Fe & 7.9 &$<$0.18 & $<$0.32\\ 
Si & 8.2 &$<$0.14 & $<$0.23\\
S & 10.4 &$<$0.27 & $<$0.41\\
Ar & 15.8 &$<$0.28 & $<$0.42\\ \hline
   \multicolumn{2}{@{}l@{}}{\hbox to 0pt{\parbox{75mm}{\footnotesize
       \par\noindent
	The emission line upper-limit at 90\% confidence level 
	(or its strength) in the cases of the abundances (of He, C, N, and O) at 0.3 and 1 solar.
       \par\noindent
     }\hss}}
\end{tabular}
\end{center}
\end{table}

\clearpage

\begin{figure}
\caption{Total band (0.5--9 keV) images in Obs1 (left, ACIS-S) and Obs2
(right, ACIS-I). The position of S1 is indicated
by the solid oval in Obs1 and by the arrow in Obs2. The dotted
lines show the background regions. The coordinate system is J2000.}
\label{fig:image}

\caption{Light curves of S1 in Obs1 ({\it left}) and Obs2 ({\it right}).
The energy bands are 0.5--9 keV({\it Total}), 0.5--2 keV({\it Soft})
and 2--9 keV({\it Hard}) from the top.
The horizontal axis is the time after the beginning of each observation.
The starting time is shown in the top of each panel. One bin is 2 ks.
The vertical axis is the detector count rate, whose scale is
normalized between the observations
by the effective area at 2 keV. The barred lines of the total band
light curves in Obs1 and Obs2 show the best-fit models by an exponential plus constant and a linear,
respectively.}
\label{fig:curve}

\caption{Time-averaged spectrum of S1 in Obs1.
The best-fit model of an absorbed thin-thermal plasma is shown with
the solid line (1T model in table \ref{tbl:1stspec}).
The arrow indicates the hump feature at 6.8 keV.}
\label{fig:spec1stmekawabs}

\caption{Time-averaged spectrum in Obs2.
The solid line shows the best-fit model of an absorbed thin-thermal plasma without edge
(1T model in table \ref{tbl:2ndspec}).
The arrow and dotted bar indicate the edge dip feature above 4 keV.}
\label{fig:spec2ndmekawabs}

\caption{Time-averaged spectrum in Obs2,  including the edge model for the best-fit model (solid line,
1T$\times$edge model in table \ref{tbl:2ndspec}). 
The details are the same as in figure \ref{fig:spec2ndmekawabs}.}
\label{fig:spec2ndmekaedgewabs}

\caption{Wide-band spectrum, showing the absorption corrected best-fit power law model in the 1st observation
and the radio fluxes in A88.
The solid line shows the best-fit radio model shown in A88.}
\label{fig:wrspow}
\end{figure}

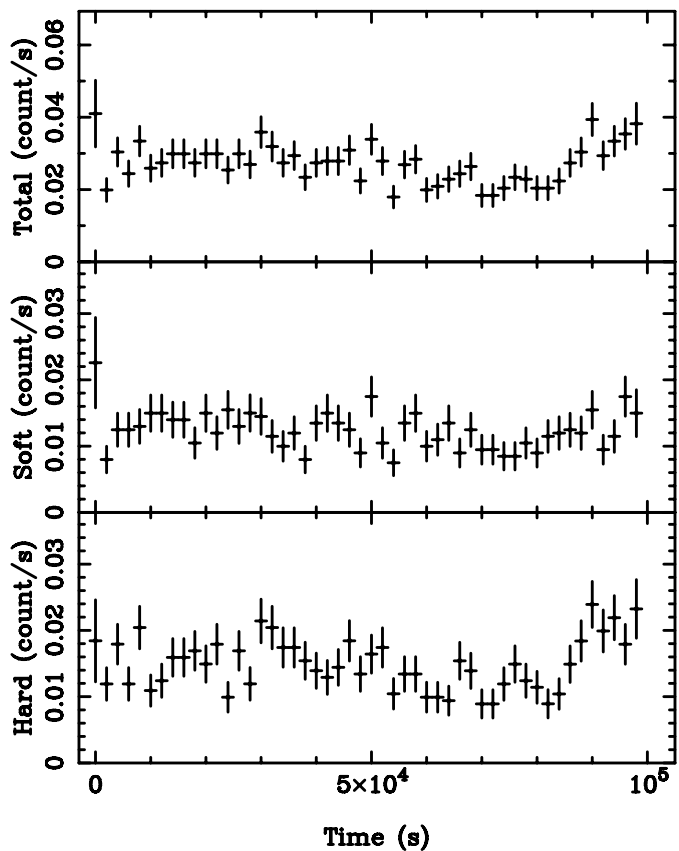
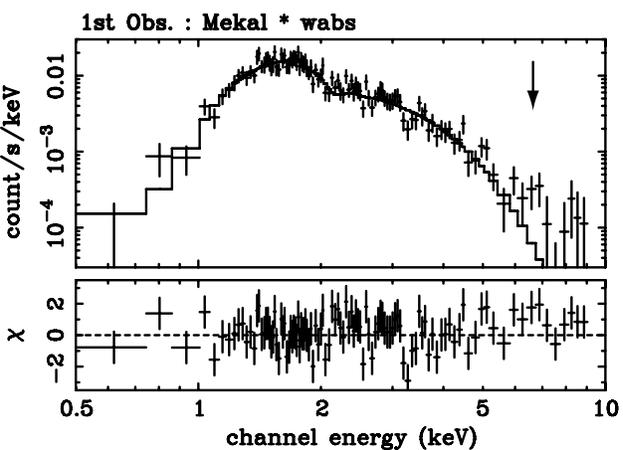
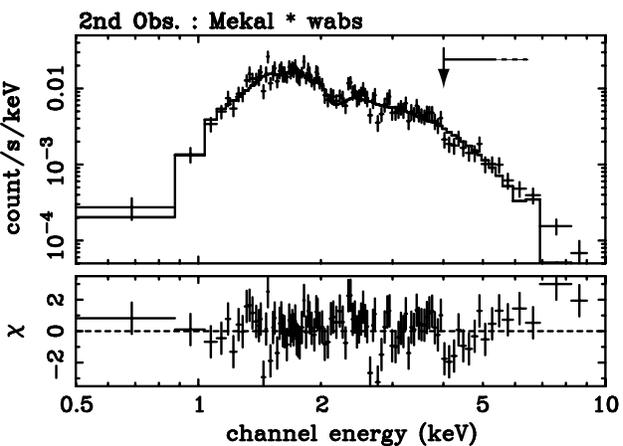
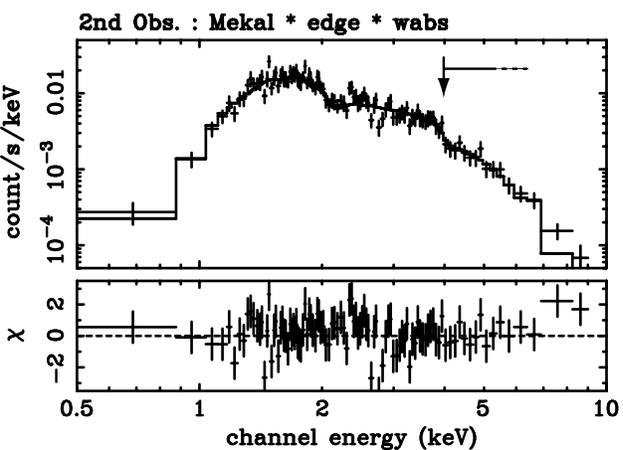
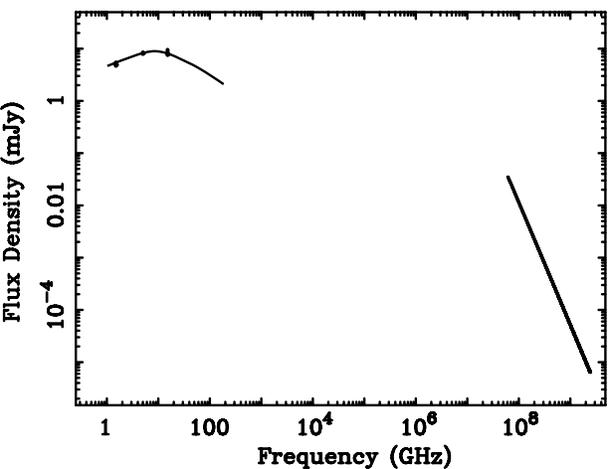
\end{document}